# Limitation to Quantum Measurements of Spacetime Distances

Y. Jack Ng[1,2,3] and H. van Dam[3]

1. School of Natural Sciences, Institute for Advanced Study, Princeton, NJ 08540
2. Center for Theoretical Physics, M.I.T., Cambridge, MA 02139
3. Institute of Field Physics, Department of Physics and Astronomy, University of North Carolina, Chapel Hill, NC 27599-3255

*Anything I have written on the topic is primarily testimony as to how important I consider it.*  — J.A. Wheeler, private correspondence.

**ABSTRACT**

Inspired by the work of Wheeler among others, we have studied the problem of quantum measurements of space-time distances by applying the general principles of quantum mechanics as well as those of general relativity. Contrary to the folklore, the minimum error in the measurement of a length is shown to be proportional to the one-third power of the length itself. This uncertainty in space-time measurements implies an uncertainty of the space-time metric and yields quantum decoherence for particles heavier than the Planck mass. There is also a corresponding minimum error in energy-momentum measurements.

We start by recalling the fundamental nature of space-time distance measurements.[1-4] In quantum mechanics one specifies a space-time point simply by its coordinates, without bothering to give a prescription as to how these coordinates are to be measured. However, general relativity ordains that coordinates do not have any meaning independent of observations; in other words, according to relativity, a coordinate system is defined only by explicitly carrying out space-time distance measurements. We will pay heed to this dictum of general relativity, and will follow Salecker and Wigner[1] to use clocks and light signals to measure distances.[5][f1]

Suppose we want to measure the length between two spatially separated points $A$ and $B$. The measurement can be carried out in the following way. A clock is put at point $A$. Set the clock to read zero when a light signal is sent from $A$ towards $B$ where a mirror is stationed to reflect the light signal back to $A$. From the reading of the clock, to be denoted by $t$, when the light signal arrives at $A$, one deduces that the length $AB$ is given by $\ell = ct/2$ where $c$ denotes the speed of light. There are two major sources of errors in the length measurement: one comes from the uncertainty principle of quantum mechanics, and the other is due to spacetime curvature effects.

First we note that the clock is not absolutely stationary, its spread in speed being given by the uncertainty principle of quantum mechanics,

$$\delta v = \frac{\delta p}{m} \gtrsim \frac{1}{2} \frac{\hbar}{m\delta\ell_{QM}}, \tag{1}$$

where $m$ is the mass of the clock. Since the clock is the agent in measuring the length, that

---

[f1] Our work has some overlap with that of Ref. 6.

it is not stationary implies an uncertainty in the length measurement given by, at time $t$,

$$\delta\ell_{QM}(t) = t\delta v \gtrsim \frac{1}{2} \frac{\hbar}{m\delta\ell_{QM}(0)} t = \frac{1}{m} \frac{\hbar}{\delta\ell_{QM}(0)} \frac{\ell}{c}, \tag{2}$$

where in the last two steps we have used Eq.(1) and $t = 2\ell/c$ respectively. Next we minimize $\delta\ell_{QM}(0) + \delta\ell_{QM}(t)$ so that the uncertainty in the length measurement due to quantum mechanical effects is given by

$$(\delta\ell_{QM})^2 \gtrsim \frac{\hbar\ell}{mc}. \tag{3}$$

But the uncertainty in the position of the clock cannot be made arbitrarily small by making the clock very massive as that would disturb the spacetime curvature.[7] If one assumes the clock to be spherically symmetric, and to have a radius (to be denoted $a$) larger than the Schwarzschild radius $r^* = \left(\frac{2Gm}{c^2}\right)$ where $G$ is the gravitational constant, $\delta\ell_c$, the error in the length measurement caused by the curvature, may be calculated from the Schwarzschild solution. The result is $r^*$ multiplied by a logarithm involving $\frac{r^*}{a}$ and $\frac{r^*}{a+\ell}$. For $a \gg r^*$ one finds $\delta\ell_c = \frac{1}{2} r^* \ln \frac{a+\ell}{a}$, hence

$$\frac{Gm}{c^2} \lesssim \delta\ell_c. \tag{4}$$

The combined error in the length measurement, $\delta\ell = \delta\ell_{QM} + \delta\ell_c$, , due to quantum mechanical and curvature effects, can be minimized by adjusting $m$. Using Eqs. (3) and (4) we find

$$\delta\ell \gtrsim (\ell\, \ell_P^2)^{1/3}, \tag{5}$$

where $\ell_P = \left(\frac{\hbar G}{c^3}\right)^{1/2}$ is the Planck length.[f2] We expect the presence of the mirror at point $B$ to contribute an error of comparable magnitude. We can also deduce the minimum error

---

[f2] Carrying out the measurement at non-zero temperature results in an additional error.

in time interval measurements by using the same experimental set-up:

$$\delta t \gtrsim (t\, t_P^2)^{1/3} ,\qquad(6)$$

where $t_P = \ell_P/c$ is the Planck time. These errors in spacetime measurements induce an uncertainty in the spacetime metric. Noting that $\delta\ell^2 = \ell^2 \delta g$ and using Eqs. (5) and (6) we readily get

$$\delta g_{\mu\nu} \gtrsim \left(\frac{\ell_p}{\ell}\right)^{2/3} \sim \left(\frac{t_p}{t}\right)^{2/3} .\qquad(7)$$

Our results (Eqs. (5)-(7)) should be contrasted with those according to the canonical[8] viewpoint. The derivation of the latter goes as follows. The vacuum functional for the theory of pure gravity has roughly the form

$$Z \sim \int \mathcal{D}[g] \exp\left\{\frac{ic^3}{\hbar G} \int \left(\frac{\partial g}{\partial x}\right)^2 d^4x\right\} .\qquad(8)$$

Thus if one is making measurements in a spacetime region of volume $\ell^4$, contributions to the Feynman integral are more or less in phase until variations in the gravitational field amplitudes from their classical values become as large as

$$\delta g \gtrsim (\hbar G/c^3)^{1/2}/\ell = l_P/\ell .\qquad(9)$$

These represent the quantum fluctuations of the gravitational fields. They give rise to errors in spacetime measurements which are constants:

$$\delta\ell \gtrsim \ell_P ;\qquad \delta t \gtrsim t_P .\qquad(10)$$

Let us see how reasonable the canonical results (Eqs.(9) and (10)) are in the light of the analysis given above for our actual experimental set-up involving a clock, a mirror, and light signals. There, to obtain the canonical results instead of Eqs. (5) - (7), all one has to do is to replace Eq. (4) by the requirement $\frac{Gm}{c^2} \lesssim \ell$. But this requirement is trivially true, for otherwise the mirror would be required to be inside the Schwarzschild radius of the clock, a nonsensical situation. One would think one could impose a more restrictive (but still physically sensible) requirement to arrive at a more useful and better bound for the minimum errors in spacetime measurements (such as those given by Eq. (5) - (7)).

Two more comments on the results Eq. (5) - (7):

(1) Mathematically it is not surprising that the uncertainty in the length $\ell$ involves the two lengths in the problem: $\ell$ and $\ell_P$. There is an analogous result which is actually relevant for long thin rulers. A quantum mechanical calculation for a 1-dimensional chain of $N$ ions with a spring of constant $k$ between successive ions gives in the high temperature limit, for the uncertainty in the length, [f3]

$$\delta\ell = \frac{1}{\pi}\sqrt{N\overline{\Delta x_i^2}} = \frac{1}{\pi}\sqrt{Nb\frac{\overline{\Delta x_i^2}}{b}} = \frac{1}{\pi}\sqrt{\ell\frac{\overline{\Delta x_i^2}}{b}},$$

where $b$ is the distance between two successive ions when the spring is relaxed, and where $\overline{\Delta x_i^2} = \frac{1}{2}\frac{k_B T}{k}$, i.e., the mean square displacement of a mass on a spring of force constant $k$. Thus, for a long thin ruler, the uncertainty in the length depends on both the length $\ell$ itself and the lattice constant $b$. Note that $\delta\ell$ is proportional to $\ell^{1/2}$. [In the zero temperature limit, one finds $\delta\ell$ to be proportional to $\sqrt{\log \frac{\ell}{b}}$.] This is one of the reasons that long thin rulers are not the best tools to measure distances. In addition, being macroscopic objects, rulers will influence other objects in the measurement process through their gravitational attraction. The Lorentz contraction of rulers will also complicate matters.

---

[f3] This result is originally due to E.P. Wigner (in response to a question raised by one of us (HvD),) private communication. Details of the calculation will appear elsewhere.

(2) Because the Planck length is so small, even for the universe [f4] for which $\ell = 10^{10}$ light-years, $\delta\ell$ is only $10^{-13}$ cm long, the size of a nucleus, which is quite small, but in principle is there.

As a simple application of Eqs. (5)-(7), we deduce the minimum error in momentum-energy measurements. Imagine sending a particle of momentum $p$ to probe a certain structure of spatial extent $\ell$ so that

$$p \sim \frac{\hbar}{\ell} \,. \tag{11}$$

Consider the coupling of the metric to the energy-momentum tensor of the particle,

$$(g_{\mu\nu} + \delta g_{\mu\nu})t^{\mu\nu} = g_{\mu\nu}(t^{\mu\nu} + \delta t^{\mu\nu}) \,, \tag{12}$$

where we have noted that the uncertainty in $g_{\mu\nu}$ can be translated into an uncertainty in $t_{\mu\nu}$. Eqs. (7) and (12) can now be employed to give

$$\delta p \gtrsim p \left(\frac{\ell_P}{\ell}\right)^{2/3} , \tag{13}$$

which, with the aid of Eq. (11), yields

$$\delta p \gtrsim p \left(\frac{p}{m_P c}\right)^{2/3} , \tag{14}$$

where $m_P = (\hbar c/G)^{1/2}$ is the Planck mass.

---

[f4] Since this conference is held in honor of Prof. J.A. Wheeler, it may not be out of place here at the mentioning of the word "universe" to quote him (private correspondence): '[I recall] the well-known statement of Rutherford, "When a student of mine uses the word "universe", I tell him it is time for him to leave." But maybe that's why so many of us live in America!'

[An alternative derivation of Eq. (14) is provided by considering $\delta p$, the "uncertainty" of the momentum operator $p = \frac{\hbar}{i} \frac{\partial}{\partial x}$, associated with $\delta x = (x\ell_P^2)^{1/3}$ (Eq. (5)). For any function $f(x)$, $(\delta p)f$ is given by

$$(\delta p)f = \frac{\hbar}{i} \left[ \delta x \, \frac{\partial^2 f}{\partial x^2} + \frac{\partial f}{\partial x} \frac{\partial \delta x}{\partial x} \right] \, .$$

Taking the function $f(x)$ to be the linear momentum eigenstate $f = \exp(ipx/\hbar)$, one gets

$$\left| (\delta p) e^{\frac{ipx}{\hbar}} \right| = \hbar \left| \left[ i\frac{p^2}{\hbar^2} \ell_P^{2/3} x^{1/3} + \frac{1}{3} \frac{p}{\hbar} \ell_P^{2/3} x^{-2/3} \right] e^{\frac{ipx}{\hbar}} \right| \, .$$

The minimum value of $|\delta p|$ is attained at $x \sim \frac{2}{3} \frac{\hbar}{p}$ yielding Eq. (14).]

The analogous statement for the minimum error in energy measurements is

$$\delta E \gtrsim E \left( \frac{E}{m_P c^2} \right)^{2/3} \, . \tag{15}$$

For energy-momentum near the Planck scale, the error is not negligible; for example, at the Grand unification scale $\sim 10^{16}$ GeV, the error is of order 1%. In analyzing a high-energy experiment an experimentalist should keep in mind that energy-momentum conservation holds only up to the errors given by Eqs. (14) and (15). In passing, we mention that the minimum errors in measurements given by Eq.(6), (7), (14) and (15) are fixed by dimensional analyses once the minimum error in spatial distance measurements is found given by Eq. (5).

As another application of the above results Eqs. (5)-(7), let us consider the quantum (de)coherence phenomenon for a scalar particle of mass $m$ moving slowly. Let us assume that the particle satisfies the Schrödinger-Klein-Gordon type equation,

$$i\hbar \frac{\partial}{\partial t} \psi = \left( -\frac{\hbar}{2m} \frac{\partial^2}{\partial x^2} + V(x,t) + g^{00} mc^2 \right) \psi \, , \tag{16}$$

where we have kept only the most important term involving $g^{\mu\nu}$. We are interested here in the effects caused by the uncertainty in the metric given by Eq. (7). Obviously $\delta g$ induces a multiplicative phase factor in the wave-function

$$\psi \to e^{i\delta\varphi}\psi \,, \tag{17}$$

where

$$\delta\varphi = \frac{1}{\hbar} \int^t mc^2 \delta g^{00} dt' \,. \tag{18}$$

For consistency the integral should be fairly insensitive to the lower integration limit as long as $t \gg t_P$. If one is making measurements in a short time interval, contributions to the phase of the wave-function from the different time elements in this time interval will be more or less in phase (i.e. $\delta\varphi \ll 1$) until the time interval reaches the decoherence time $t_D$ when $\delta\varphi$ becomes sizable, i.e.,

$$1 \sim \frac{1}{\hbar} \int_0^{t_D} mc^2 \left(\frac{t_P}{t'}\right)^{2/3} dt' \sim \frac{mc^2}{\hbar} t_P^{2/3} t_D^{1/3} = \ell_P^{2/3} \ell_D^{1/3}/\lambda_c \,, \tag{19}$$

where $\lambda_c = \frac{\hbar}{mc}$ is the Compton wave-length of the particle of mass $m$, and $\ell_D = ct_D$ is what we will call the decoherence length. The system can be treated classically if the decoherence length is less than the Compton wave-length, in other words, if

$$\ell_D \leq \lambda_c \,,$$

or, via Eq. (19),

$$m \gtrsim m_P \,. \tag{20}$$

Therefore, due to the uncertainty of the space-time metric, it suffices to give a particle heavier than the Planck mass a classical treatment.

We thank Prof. J.A. Wheeler for an enjoyable correspondence, and G. Domokos for calling our attention to Ref. 6. The first author thanks the faculty (especially S. Adler) at the School of Natural Sciences, Institute for Advanced Study and the faculty (especially A. Guth) at the Center for Theoretical Physics, M.I.T. for kind hospitality. This work was supported in part by the U.S. Department of Energy under Grant No.DE-FG05-85ER40219 and by the Z. Smith Reynolds Fund of the University of North Carolina.